# Tunable plasmons in large area WTe$_2$ thin films

Chong Wang[1,2], Yangye Sun[3], Shenyang Huang[1,2], Qiaoxia Xing[1,2], Guowei Zhang[4], Chaoyu Song[1,2], Fanjie Wang[1,2], Yuangang Xie[1,2], Yuchen Lei[1,2], Zhengzong Sun[3*], Hugen Yan[1,2*]

[1]State Key Laboratory of Surface Physics and Department of Physics, Fudan University, Shanghai 200433, China.

[2]Key Laboratory of Micro and Nano-Photonic Structures (Ministry of Education), Fudan University, Shanghai 200433, China.

[3]School of Microelectronics and State Key Laboratory of ASIC and System, Fudan University, Shanghai 200433, China

[4]Institute of Flexible Electronics, Northwestern Polytechnical University, Xi'an 710072, China

*E-mail: hgyan@fudan.edu.cn (H. Y.); zhengzong_sun@fudan.edu.cn (Z. S.)

**Abstract:**

The observation of the electrically tunable and highly confined plasmons in graphene has stimulated the exploration of interesting properties of plasmons in other two dimensional materials. Recently, hyperbolic plasmon resonance modes are observed in exfoliated $WTe_2$ films, a type-II Weyl semimetal with layered structure, providing a platform for the assembly of plasmons with hyperbolicity and exotic topological properties. However, the plasmon modes were observed in relatively thick and small-area films, which restrict the tunability and application for plasmons. Here, large-area (~ cm) $WTe_2$ films with different thickness are grown by chemical vapor deposition method, in which plasmon resonance modes are observed in films with different thickness down to about 8 nm. Hybridization of plasmon and surface polar phonons of the substrate is revealed by mapping the plasmon dispersion. The plasmon frequency is demonstrated to be tunable by changing the temperature and film thickness. Our results facilitate the development of a tunable and scalable $WTe_2$ plasmonic system for revealing topological properties and towards various applications in sensing, imaging and light modulation.

# I. INTRODUCTION

Two dimensional (2D) plasmons have attained much attention since the discoveries of tunable graphene plasmons with strong energy confinement capability [1, 2], distinct carrier density dependence [3] and strong coupling to polar phonons of substrate [4, 5] and vibrations of adsorbed molecules [6]. With the development of the family of 2D materials, it is desirable to reveal new realizations of 2D plasmons beyond graphene [7-9]. One direction is to probe the plasmons in three-dimensional (3D) gapless topological semimetals, 3D analogs of graphene [10], such as Dirac plasmons discovered in type-II Dirac semimetal $PtTe_2$ and topologically nontrivial interband plasmons in type-II Weyl semimetal $MoTe_2$ by electron energy loss spectroscopy (EELS) [11, 12]. Especially, Weyl semimetals are one kind of topological semimetals with pairs of Weyl nodes displaying opposite chirality, where the collective modes have been predicted to exhibit a number of unusual responses, such as exotic plasmon dispersion due to chiral anomaly [13], nonreciprocal surface plasmons in Weyl semimetal with broken time-reversal symmetry [14], Fermi arc plasmons with hyperbolic plasmon dispersion [15, 16]. In addition, the temperature dependence of the plasmon frequency gives some indication on the topological phase transition [12, 17]. Another direction is to examine layered materials with in-plane anisotropic electrodynamic responses, which have been predicted to potentially realize plasmons with hyperbolic dispersion [18]. In contrast to the hyperbolic metamaterials and metasurfaces with artificial structures, such natural hyperbolic surface based on 2D materials sustain tunable hyperbolic plasmons with larger field confinement and higher photonic density of states.

The type-II Weyl semimetal $WTe_2$ is one of the topological semimetals with layered structures and in-plane anisotropy [19-22]. Recently, plasmons with hyperbolic isofrequency contours were demonstrated in this natural 2D material surface in the far-IR range, making $WTe_2$ an ideal platform to explore both the topological and hyperbolic properties of plasmons [17]. However, limited by the exfoliation method, the $WTe_2$ plasmons were only observed in relatively thick (above 50 nm) and small-area (about 300 μm) films, in which it is difficult for carrier density tuning and the scalable application of plasmons in the far-IR range. Since the Weyl points in undoped $WTe_2$ is well above the Fermi energy, it is essential to tune the carrier density in order to probe the Weyl physics [20]. Electrical tuning is also important for the propagation-manipulation of hyperbolic plasmon polaritons [9, 18]. Moreover, a serious of exotic electrical properties, e.g. two dimensional topological insulator [23], quantum spin Hall effect up to 100 K [24] and electrical tunable superconductivity [25, 26] and ferroelectricity [27, 28] have been demonstrated in monolayer or few layer $WTe_2$ films. It is appealing to interrogate these exotic properties for plasmons if we can observe plasmons in thin films with large area.

Recently, large area $WTe_2$ films with different thickness are reported to be grown by chemical vapor deposition (CVD) method [29-31]. In our study, plasmon resonance modes based on centimeter-scale CVD-grown $WTe_2$ films with different thickness are investigated using the Fourier transform infrared spectroscopy (FTIR). Hybridization with surface polar phonon of substrates and temperature- and thickness-dependence of plasmons are revealed, setting a foundation for further study of exotic properties and applications of 2D plasmons in WTe$_2$.

## II. RESULTS AND DISCUSSION

A schematic of the growth method for WTe$_2$ film is shown in Figure 1a. W seed layers with different thickness were deposited on Si and Si/SiO$_2$ (300 nm SiO$_2$) substrates by a magnetron sputtering system (DE500). Then the W films were oxidized in air (400 °C, 15 min) to make surfaces smoother. The obtained WO$_x$ films and Te powders (about 300 mg) were placed in a quartz crucible. The CVD growth of WTe$_2$ films was performed by a two-zone furnace (OTF-1200X). WO$_x$ films were tellurized into WTe$_2$ films after annealing in Te vapor at 600 °C for 30 min under ambient pressure with argon and hydrogen (30 sccm Ar, 40 sccm H$_2$) as the carrier gas. The thickness of the W film determines the final thickness of the WTe$_2$ film.

We firstly study the WTe$_2$ films with the thickness of 30 nm on Si and Si/SiO$_2$ substrates. The inset of Figure 1b shows the optical image of WTe$_2$ film on Si/SiO$_2$ substrate, exhibiting a uniform and smooth surface. The Raman spectrum (Figure 1b) is consistent with that in bulk crystals, confirming the T$_d$ phase of the film [32]. To characterize the optical properties of the grown WTe$_2$ film, extinction spectra were measured as illustrated in Figure 1a, with $T$ and $T_0$ being the transmission through the sample and bare substrate, respectively. The transmission spectra measurements were performed by a Bruker FTIR spectrometer (Bruker Vertex 70v) integrated with a liquid-helium-cooled silicon bolometer as a detector. The extinction spectrum of WTe$_2$ film on Si substrate is shown in Figure 1c [33], which consists of two components: the intraband transition component (Drude response) which dominates the low frequency regime, and the interband transition component which has larger weight at higher frequency. Similar extinction spectra can be also observed for the film on Si/SiO$_2$ substrate (inset of Figure 1c), except of a dip at around 460 cm$^{-1}$ which is the same frequency as the transverse optical (TO) phonon on SiO$_2$ [4]. This dip is originated from the different absorption of the TO phonon for substrates with and without WTe$_2$ films. To probe the plasmon resonance modes, ribbon arrays (insets of Figures 1d and 1e) are fabricated on the WTe$_2$ films by electron-beam lithography and subsequent reactive ion etching with sulphur hexafluoride (SF$_6$) as the reaction gas. As shown in Figures 1d and 1e, clear plasmon resonance peaks can be observed in the extinction spectra with the light polarization perpendicular to the ribbons. The frequency and intensity differences for those two cases come from the different dielectric constants of the substrates.

Because of the distinct Coulomb interaction in 2D materials, the 2D plasmon frequency has a general $\sqrt{q}$ dependence when only considering free carriers [34]. For polar substrates like SiO$_2$, where surface polar phonons exist, softened plasmon dispersion and enhanced transparency at phonon energy will be induced due to the plasmon-phonon coupling, which can be employed for chemical sensing [35]. Here we fabricated a set of ribbon arrays of WTe$_2$ films on the two kinds of substrates to investigate the plasmon resonance modes as a function of ribbon width, as shown in Figures 2a and 2b. Based on the large film area, the measured spectral range can be extended to the Terahertz (THz) range (down to about 30 cm$^{-1}$ or 1 THz). The plasmon frequencies extracted from Figures 2a and 2b are displayed in Figures 2c and 2d as a function of wave vector $q$ determined by $\pi/W$, where $W$ is the width of the ribbon. As shown in Figure 2c, the plasmon dispersion on Si substrates can be well described by $\sqrt{q}$ relation (the red line). The plasmon dispersion on Si/SiO$_2$ substrates (Figure 2d) follows the $\sqrt{q}$ relation (the blue dashes line) at lower frequency, but departs from it at higher frequency, which is originated from the plasmon-phonon coupling [4, 5]. Random phase approximation considering the Fröhlich coupling of plasmon with surface optical phonon is used to fit the plasmon dispersion on Si/SiO$_2$ substrates [4, 33, 36]. The fitting result is plotted as the blue line in Figure 2d, which has excellent agreement with the experimentally observed plasmon frequencies (blue circles). The fitted plasmon resonance width on both substrates are summarized in the inset of Figure 2d. In contrast to the constantly large values on Si substrates (red squares), the resonance width on Si/SiO$_2$ substrates (blue circles) dramatically decreases at small ribbon width, which is attributed to the decreased damping rate due to plasmon-phonon hybridization, since phonons have much longer lifetime [4]. A dip at around 500 cm$^{-1}$, which has the same frequency as the position of the longitudinal optical (LO) phonon of the SiO$_2$ film, can be observed in the extinction spectra of Figure 2b and Figure 1e. This is a result of the plasmon-phonon coupling, which has been observed in previous work [5]. Thus, the dispersion softening and the resonance width reduction as well as the dip at the LO phonon, all indicate the strong hybridization between the plasmon and the surface polar phonons on Si/SiO$_2$ substrates.

In the previous optical measurement on the bulk WTe$_2$ and the exfoliated WTe$_2$ films, the frequency of the bulk plasmons and the 2D plasmons both exhibit temperature dependence [17, 21]. However, only the in-plane principle axes were investigated. The CVD-grown films here are poly-crystals with a combination of all crystal axes, so it is interesting to check if the plasmons in the CVD films have the same temperature dependence as in the single crystal films. Figure 3a shows the temperature evolution of the plasmon resonance mode in a 2 μm-width ribbon array of the CVD-grown $WTe_2$ film on the $Si/SiO_2$ substrate. As shown in Figure 3b, the plasmon frequency exhibits a blue shift upon heating, with an increase rate of about 75% between 300 K and 10 K, revealing a strong tunability by temperature. The plasmon intensity (Figure 3c) also increases as temperature rises. The enhancement of the frequency and intensity stems from the increased carrier density [37] and possibly decreased effective mass at high temperature which give rise to increased Drude weight upon heating. The intensity decreases above 200 K because of the transfer of the plasmon weight to the high energy hybrid mode due to the plasmon-phonon hybridization. It is noted that the previous reported plasmon frequencies in single crystal films remain almost unchanged below 150 K [17], while the plasmon frequency in the CVD films keeps decreasing all the time as the temperature decreases. The difference might be attributed to the possible doping in the CVD growth. The plasmon width is as well tunable by temperature, revealed by a sharper plasmon mode at 10 K as shown in Figure 3d. The resonance width has a reverse temperature-dependence above 200 K, which is originated from the hybridization as the plasmon frequency approaches that of the surface polar phonon.

In 2D material, the Drude weight and plasmon frequency are both determined by the sheet conductivity which is proportional to the film thickness if the electronic structures are unchanged. CVD-grown films provide a convenient way to tune the plasmon frequency by controlling the film thickness. Figure 4b shows the image of WTe$_2$ films with different thickness on Si/SiO$_2$ substrates. The film color changes from purple (near the color of the substrate) for 4 nm film to light yellow for 45 nm film. All films have the same Raman phonon modes as that of bulk crystal [33], demonstrating the existence of WTe$_2$ phase even in 4 nm film. Figure 4a shows the extinction spectra of the WTe$_2$ films with different thickness, where the spectra weight decreases with reducing film thickness. The fitted Drude weight is plotted in Figure 4c as a function of the film thickness. A linear behavior can be observed, consistent with the fact that the Drude weight of films are proportional to the sheet conductivity. To investigate the thickness dependence of plasmon, ribbon arrays with the same width are fabricated on films with different thickness. Figure 4d shows the plasmon spectra in films with different thickness. The plasmon resonance peak has a red shift with decreasing film thickness because of the reducing sheet conductivity. The resonance width becomes broader for thin films, presumably due to the increased impurity scattering from the increased nanoscale grains, grain boundaries, and defects in thin films [29]. From Figure 4d, we can still see clear plasmon resonance peak in 8 nm-thick film, in which thickness there have been experiments reporting successful carrier density tuning by electrical gating [38], suggesting the potential for electrostatic tuning of plasmon frequency. The measured plasmon frequency is plotted in Figure 4e as a function of film thickness. For plasmons with frequency not too close to the surface polar phonon of the substrate, the resonance frequency scales as $\sqrt{D}$ where $D$ is the Drude weight, hence is proportional to square root of the film thickness. Such scaling fits the data well, as shown in Figure 4e. It is noted that as the film thickness decreases, the spatial confinement will modify the band structure of $WTe_2$, reducing the overlap between the electron and hole pockets and decreasing the sheet carrier density [39]. However, in Figure 4e, there is no clear deviation from the square root relation for plasmons. That may be due to the larger doping effect in thinner films from defects by CVD growth, which compensates the finite-size effects on the band structure.

## III. CONCLUSION

In conclusion, temperature dependent 2D plasmons and plasmon-phonon hybridization are demonstrated in large area $WTe_2$ films grown by CVD methods with different thickness. Compared to the topological plasmons observed in bulk crystals by EELS [11, 12], the realization of plasmons in thin films by FTIR enables the turning through electrostatic gating, strain and thickness, which provides a suitable platform for studying topological plasmons, chemical sensing and applications in optoelectronics. The films studied here are isotropic poly-crystals, which cannot be used to investigate the hyperbolicity. Recently, single crystal $WTe_2$ films have been reported to be grown by CVD methods [31], whose size, however, is limited to a few hundred micrometers. By improving the size of the single crystal films grown by CVD method, it is interesting to check the hyperbolicity of plasmons in few layer $WTe_2$. Moreover, the CVD method can be used in growing other topological semimetals, like $MoTe_2$, $ZrTe_5$ and $PtTe_2$, providing a general method for studying the topological plasmon.

**Acknowledgements**

H.Y. is grateful to the financial support from the National Natural Science Foundation of China (Grant Nos. 12074085, 11734007), the National Key Research and Development Program of China (Grant Nos. 2016YFA0203900 and 2017YFA0303504), Strategic Priority Research Program of Chinese Academy of Sciences (XDB30000000)，Natural Science Foundation of Shanghai (20JC1414601). C.W. is grateful to the financial support from the National Natural Science Foundation of China (Grant No. 11704075) and China Postdoctoral Science Foundation. Z.S. is grateful to the financial support the National Natural Science Foundation of China (Grant No. 21771040), the National Key Research and Development Program of China (Grant Nos. 2016YFA0203900, 2017YFA0207303). G.Z. acknowledges the financial support from the National Natural Science Foundation of China (Grant No. 11804398), Natural Science Basic Research Program of Shaanxi (Grant No. 2020JQ-105) and Key Research and Development Program of Shaanxi (Grant No. 2020GXLH-Z-026). Part of the experimental work was carried out in Fudan Nanofabrication Lab.

Part of the experimental work was carried out in Fudan Nanofabrication Lab.

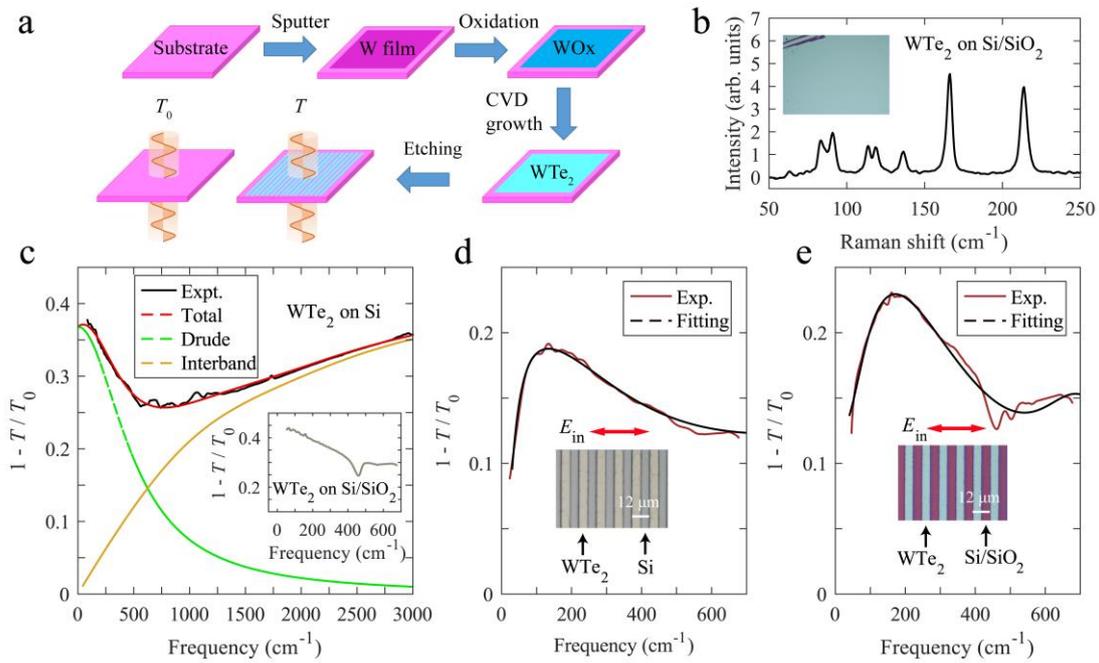

**FIG. 1.** a) Schematic of the procedure for film growth and ribbon fabrication. b) Raman spectra of the CVD-grown WTe$_2$ films. Inset: Optical microscopy image. c) Extinction spectra of CVD-grown WTe$_2$ film on Si and Si/SiO$_2$ (inset) substrates. d) and e) Extinction spectra of ribbon arrays fabricated on WTe$_2$ films on Si and Si/SiO$_2$ substrates with incident light polarized perpendicular to the ribbon arrays. Insets show the optical images of the ribbon arrays. All the films above have thickness of about 30 nm.

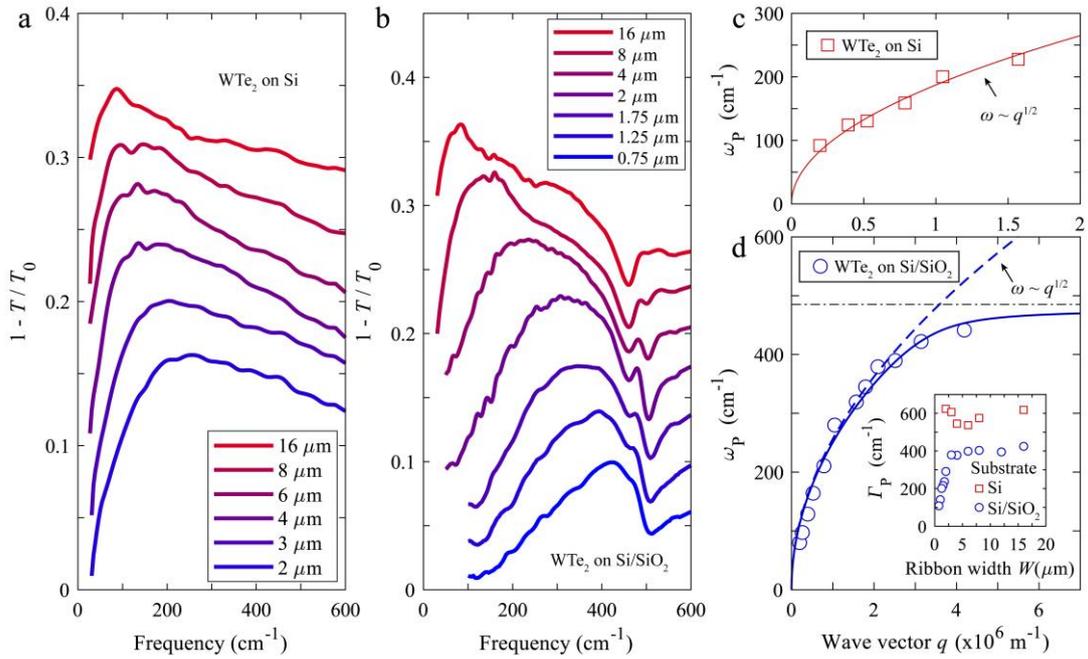

**FIG. 2.** a) and b) Extinction spectra of the plasmon resonance modes on Si and Si/SiO$_2$ substrates with different ribbon width (spectra shifted vertically for clarity). c) and d) Plasmon dispersion for WTe$_2$ films on Si and Si/SiO$_2$ substrates. The horizontal dashed line in (d) indicates the frequency of the surface optical phonon of SiO$_2$ film. Inset of (d) shows the plasmon resonance width for WTe$_2$ films on Si (rectangle squares) and Si/SiO$_2$ (blue circles) substrates as a function of ribbon width. All the films above have thickness of 30 nm.

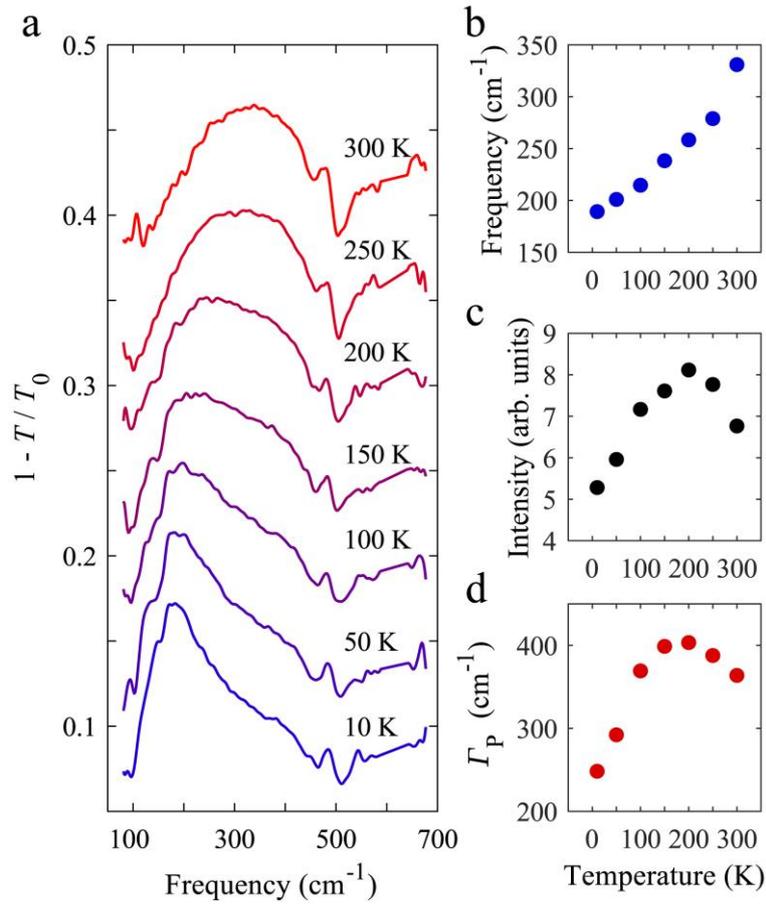

**FIG. 3.** a) Temperature dependence of the extinction spectra in the ribbon array with light polarization perpendicular to the ribbon. The ribbon width is 2 μm. b) - d) The fitted intensity, frequency and resonance width of the plasmon as a function of temperature. Film thickness is 30 nm.

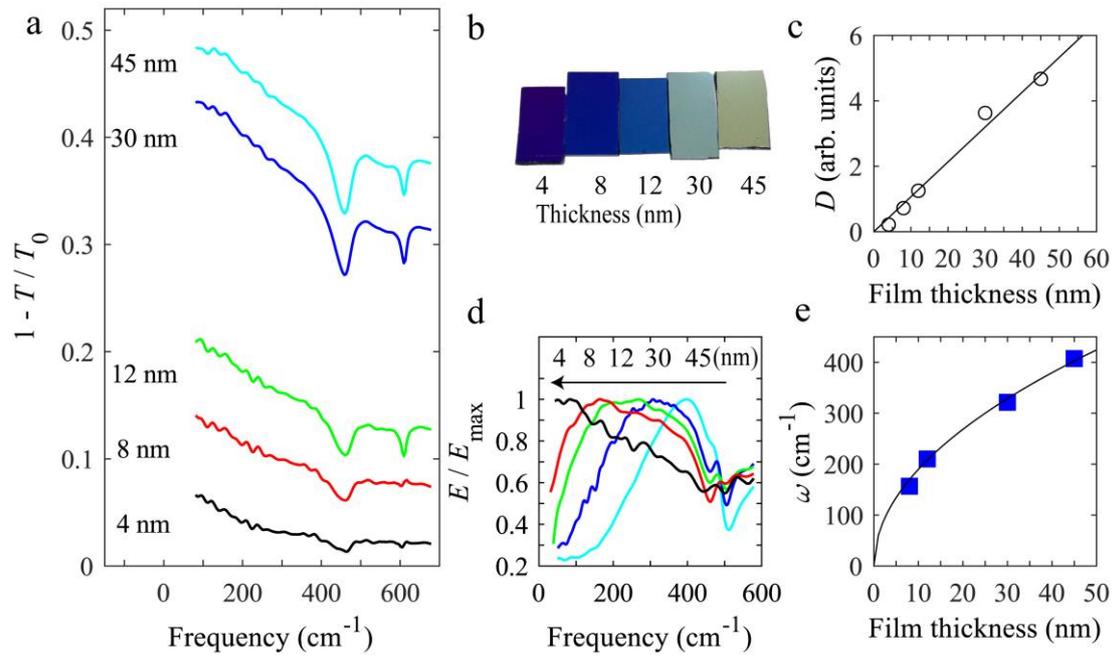

**FIG. 4.** a) Extinction spectra of CVD-grown WTe$_2$ films with different thickness. The dips at 600 cm$^{-1}$ are artifacts due to the low intensity of incident light at that frequency. (b) The optical images of WTe$_2$ films with different thickness. c) Thickness dependence of the fitted Drude response. d) Extinction spectra of ribbon arrays with 2 μm width in films with different thickness. Light has a perpendicular polarization with respect to the ribbon. e) Fitted plasmon frequency as a function of the film thickness. The black line is a square root scaling.